\documentclass[fleqn]{mnras}

\usepackage[T1]{fontenc}

\usepackage{graphicx, amsmath,natbib}

\usepackage{amssymb}

\usepackage{newtxtext,newtxmath}
\vbadness=\maxdimen
\newcommand{\concsection}[1]{ \vspace{3mm}
\noindent \textbf{\textit{{#1}}:}}
\renewcommand\textbf[1]{\textcolor{black}{#1}}
\title[MHD simulations of Abell 2146]{Turbulent magnetic fields in merging clusters: A case study of Abell 2146}

\author[Chadayammuri et al.]{
Urmila Chadayammuri,$^{1, 2}$\thanks{E-mail: urmila.chadayammuri@cfa.harvard.edu}
John ZuHone,$^{1}$
Paul Nulsen,$^{1}$
Daisuke Nagai,$^{2,3}$
Helen Russell$^{4}$
\\
$^{1}$Centre for Astrophysics | Harvard and Smithsonian, 60 Garden Street, Cambridge, MA 02143, USA\\
$^{2}$Department of Astronomy, Yale University, New Haven, CT 06520, USA\\
$^{3}$Department of Physics, Yale University, New Haven, CT 06520, USA\\
$^{4}$School of Physics and Astronomy, University of Nottingham, Nottingham NG7 2RD, UK
}

\date{Accepted XXX. Received YYY; in original form ZZZ}

\pubyear{2021}

\begin{document}
\label{firstpage}
\pagerange{\pageref{firstpage}--\pageref{lastpage}}
\maketitle

\begin{abstract}
Kelvin-Helmholtz Instabilities (KHI) along contact discontinuities in galaxy clusters have been used to constrain the strength of magnetic fields in galaxy clusters, following the assumption that, as magnetic field lines drape around the interface between the cold and hot phases, their magnetic tension resists the growth of perturbations. This has been observed in simulations of rigid objects moving through magnetised media and sloshing galaxy clusters, and then applied in interpreting observations of merger cold fronts. Using a suite of MHD simulations of binary cluster mergers, we show that even magnetic field strengths stronger than yet observed ($\beta = P_{\rm th}/P_B = 50$) show visible KHI features. This is because our initial magnetic field is tangled, producing Alfven waves and associated velocity fluctuations in the ICM; stronger initial fields therefore seed larger fluctuations, so that even a reduced growth rate due to magnetic tension produces a significant KHI. The net result is that a stronger initial magnetic field produces more dramatic fluctuations in surface brightness and temperature, not the other way around. We show that this is hard to distinguish from the evolution of turbulent perturbations of the same initial magnitude. Therefore, in order to  use observations of KHI in the ICM to infer magnetic field strengths by comparing to idealized simulations, the perturbations which seed the KHI must be well-understood and (if possible) carefully controlled. 
\end{abstract}
\begin{keywords}
clusters: theory --- clusters: simulation --- cosmological simulations:mhd --- ICM
\end{keywords}
\section{Introduction}

Merging galaxy clusters provide unique constraints on the nature of dark matter and the plasma physics of the X-ray emitting intracluster medium (ICM). This hot, diffuse gas consists of baryons trapped in the cluster gravitational potential early in its formation. It evolves with time as the cluster accretes material from the cosmic web and merges with other clusters. Internally, the ICM is also affected by radiative cooling, turbulence, and feedback from AGN in the cluster galaxies; over time, these shape its temperature, density and metallicity profiles. How energy and metals are distributed in the presence of these processes further depends on transport processes, namely viscosity, thermal conduction, and ion diffusion. \citet{Spitzer1952} and \citet[][]{Braginskii1958} derived the viscosity and thermal conductivity for a weakly collisional plasma, including how this is suppressed in the presence of a uniform magnetic field \citep{Sarazin1988}. Since then, a number of works have shown that plasma instabilities further impede cluster transport processes \citep[e.g.,][]{Schekochihin2008,Kunz2011, Kunz2012, Roberg2016}.

Magnetic fields in clusters are understood to have grown from primordial seeds of $\sim 1 nG$ \citep{Ruzmaikin1989, Subramanian2006} to the observed present-day strengths of several microgauss ($\mu G$) through turbulence and bulk flows inherent in the process of hierarchical, merger-driven structure formation \citep[e.g.,][]{Dolag2002, Medvedev2006, Vazza2014}.\textbf{Other studies suggest that primordial seeds alone fail to produce the magnetic structures in clusters today, and that additional contributions are required from star formation and AGN activity in galaxies \citep[e.g.,][]{Carilli2002, Govoni2004,Donnert2018,Xu2009,Roh2019}.} In galaxies and clusters, magnetic fields are crucial to understanding the transport of energy, metals and cosmic rays.

Most commonly, cluster magnetic fields are detected using radio observations. Using assumptions about the properties of cosmic ray electrons, we can infer the presence of diffuse magnetic fields from radio halos, radio mini-halos, and radio relics \citep[e.g.,][]{Ensslin1998,Carilli2002,Ferrari2008,Feretti2012}. Another signature is the Faraday rotation induced in background radio sources, i.e., the change in the polarisation angle of background light by magnetic fields. The polarisation angle of a linearly polarised radio source varies with wavelength, $\lambda$, as ${\rm RM}\times \lambda^2$, where the Faraday rotation measure $RM \propto \int n_e B_\parallel dl$. This signal cannot be produced by anything other than a magnetic field; the only required corrections are for the foreground magnetic field of the Milky Way, and for Faraday rotation intrinsic to the radio source (although see \citet{Johnson2020} for inherent uncertainties in the technique). The most detailed RM analysis in galaxy clusters to date have been of the Coma cluster, using seven radio sources at different radii \textbf{\citep{Bonafede2013}, and of the merging cluster Abell 2345 \citep{Stuardi2021}, with five background radio sources and two radio relics. Assuming each point source to be representative of its radial annulus, and that the field followed a radial power law $\rm RM \propto r^{-\eta}$, the Coma study found a slope $ 0.4 < \eta < 0.7$ within $1 \sigma$; in Abell 2345, the magnetic field was found to vary linearly with the local electron number density. Neither study gave results consistent with a uniform cluster-scale magnetic field.}

Such analyses, of course, come with caveats. First, the measurements are local, restricted to regions with bright background radio sources. Next, the power law form assumed for the magnetic field profile is likely too simple. Since magnetic flux is frozen into the cluster gas, the mean magnetic pressure is expected to scale with the mean turbulent pressure \citep[e.g.][]{Vijayaraghavan2017}. Further assuming that the turbulent Mach number is uniform through the plasma, the magnetic pressure can then be related to the thermal pressure $P_{\rm th} = kT\times n_g$. Observations, such as \citet[][]{Dolag2001,Govoni2017, Stuardi2021}, have indeed found the correlations between the magnetic and thermal pressure in cluster plasmas. The thermal pressure profile has been shown to have a universal form, with different slopes in the core and outskirts \citep{Nagai2007, Arnaud2010}. Constraining more parameters inevitably requires sampling the diffuse field at more points. Further, the RM tells us only about the component of the magnetic field along our line-of-sight. Converting this into the total field strength usually entails the assumption that its distribution is isotropic. This is certainly not the case during a cluster merger, where the field lines drape around the dense, low-entropy core of the subcluster \citep[e.g.,][]{Dursi2008, Lyutikov2006}. This list is not meant to be exhaustive, but illustrative of the limitations of a single method of measurement, and to emphasise the need for alternative metrics of the cluster field. While none of them can be perfect in isolation, each can provide more local measurements, which can compose a fuller picture of the global magnetic field. 

Given the density and temperature of the ICM, the mean free path of the electrons and ions are on the order of kpc, whereas given the magnetic field strength of $\sim\mu$G and electron and ion temperatures, their Larmor radii are no more that the solar radius. This means that the transport across the field lines is heavily suppressed, making the transport processes highly anisotropic \citep{Ruszkowski2010, Kunz2011, Kunz2011b, Kunz2012, Santos2014}. If the field lines are highly tangled, transport may be suppressed generally.  Further suppression can result from plasma instabilities, which may drive waves that scatter the ions or electrons strongly \citep[e.g.,][]{Schekochihin2008, Roberg2016}. Because our understanding of the net effect of these instabilities and anisotropies is still evolving, extracting field strengths using constraints on the effectiveness of transport processes still entails significant theoretical uncertainty.

Yet another measure of cluster magnetic fields uses observed X-ray features of merging clusters. As the dense, low-entropy core of a subcluster moves through the hotter surrounding ICM during a merger, magnetic fields drape around its leading edge, forming a highly magnetized layer parallel to the front surface \citep{Lyutikov2006}; this draping also changes the geometry of the cold front, leaving it with smaller opening angles \citep[e.g.,][]{Dursi2008, ZuHone2011}. Noting that such a sheath inhibits the growth of perturbations, \citet{Vikhlinin2001} used the lack of observed Kelvin-Helmholtz Instabilities (KHI) at the leading edge of the cold front in the merging cluster Abell 3667 to estimate $B \sim 10\mu G$. Similarly, if the KHI in sloshing cold fronts in Virgo are suppressed by ICM viscosity, \citet{Roediger2013a} found that this viscosity would have to be at least $\lesssim 0.1\nu_{\rm Spitzer}$. Magnetic fields offer one channel for such a viscosity suppression. 

\begin{figure}
    \centering
    \includegraphics[width=0.45\textwidth]{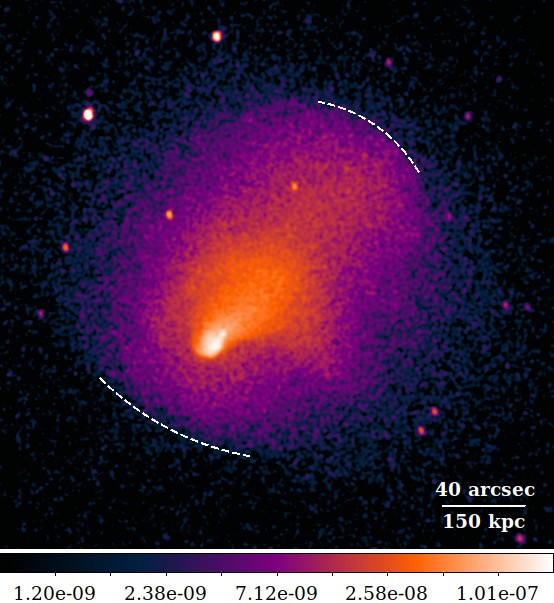}
    \caption{\textit{Chandra }surface brightness maps of Abell 2146. The image shows photons from 0.5-2.0 keV collected over 400 ks, and is described further in \citet{Russell2011}.}
    \label{fig:obs}
\end{figure}

Here, we present just such a study of the merging galaxy cluster Abell 2146, whose complex ICM was first observed in the X-ray with the \textit{Chandra} X-ray Observatory \citep{Russell2010}. This observation revealed some of the first merger shocks detected since the Bullet Cluster. Being less massive and thus cooler than the Bullet Cluster, the gas in Abell 2146 is better suited for observations in the Chandra band, so that surface brightness and temperature maps can be mapped in unprecedented detail. Fig \ref{fig:obs}. These various plasma processes, in turn, depend on the magnetic field within the ICM. In this paper, we describe the observable consequences of magnetic fields of different strengths on X-ray observations of merging clusters similar to Abell 2146. 

The X-ray features are primarily determined by the mass profiles of the subclusters, their relative velocity, and the geometry of the merger. These parameters were constrained through an extensive parameter study in \citet{Chadayammuri2022} (henceforth Paper I). In this paper, we explore the role of adding magnetic fields with properties expected from observations. We make predictions not only for X-ray maps, but also for the Faraday rotation. Section \ref{sec:sims} describes the simulation setup. We show our results for the effects of mergers on the magnetic field, and of the magnetic field on observables, in Section \ref{sec:results}. We discuss caveats and future work in Section \ref{sec:discuss}, and wrap up with conclusions in Section \ref{sec:conclusions}.

\section{The Simulation Setup}
\label{sec:sims}
We run a suite of idealized \textbf{binary merger} simulations with a GPU-accelerated Adaptive MEsh Refinement code, GAMER-2 \citep{Schive2018}, \textbf{which in its latest version includes a magnetohydrodynamics (MHD) solver \citep{Zhang2018}.}

\textbf{We solve the following MHD equations (in conservative form, and written here in Gaussian units):}

\begin{equation}
\frac{\partial \rho_g}{\partial t} + \nabla\cdot(\rho_g\mathbf{v})=0
\end{equation}
\begin{equation}
\frac{\partial(\rho_g \mathbf{v})}{\partial t} + \nabla\cdot\left(\rho_g\mathbf{v}\mathbf{v}-\frac{\mathbf{B}\mathbf{B}}{4\pi}\right)+\nabla p=\rho_g \mathbf{g} 
\end{equation}
\begin{equation}
\frac{\partial E}{\partial t} + \nabla\cdot\left[\mathbf{v}(E+p)-\frac{\mathbf{B}(\mathbf{v}\cdot\mathbf{B})}{4\pi}\right]=\rho_g\mathbf{g}\cdot\mathbf{v} 
\end{equation}
\begin{equation}\label{eqn:induction}
\frac{\partial \mathbf{B}}{\partial t} + \nabla\cdot(\mathbf{v}\mathbf{B}-\mathbf{B}\mathbf{v})=0,
\end{equation}
\textbf{where $\rho_g$ is the gas density, $\textbf{v}$ is the gas velocity, and $\textbf{B}$ is the magnetic field strength. The total energy density $E$, total pressure $p$, and gravitational acceleration $\textbf{g}$ have the usual definitions:}
\begin{equation}
p = p_{\text{th}}+\frac{B^2}{8\pi}
\end{equation}
\begin{equation}
E = \frac{1}{2}\rho v^2+\epsilon+\frac{B^2}{8\pi}
\end{equation}
\begin{equation}
\mathbf{g}=-\nabla\phi
\end{equation}
\textbf{where $\epsilon$ is the gas internal energy per unit volume, and the gravitational potential $\phi$ is solved using Poisson's equation:}
\begin{equation}
\nabla^2\phi=4\pi G(\rho_g+\rho_{\text{DM}})
\end{equation}
\textbf{where $\rho_{\rm DM}$ is the dark matter density. For the gas, we assume an ideal gas equation of state with $\gamma=5/3$.}

\textbf{The equations of MHD are solved numerically using  a finite-volume, high-order Godunov scheme combined with a constrained transport method, similar to that employed in other codes \citep[][]{Fromang2006,Stone2008,Stone2009,Lee2009,Mignone2012,Bryan2014}, which guarantees that the evolved magnetic
field satisfies the divergence-free condition by evolving the induction equation \citep{Evans1988}, preserving this condition to machine precision. In our simulations,
the order of the Riemann solver for the hydrodynamic fluxes corresponds to the Piecewise-Parabolic Method of \citet{Colella1984}, which is ideally suited for capturing shocks and contact
discontinuties (such as the cold fronts that appear in our simulations). GAMER-2
also includes an $N$-body module which uses the particle-mesh method to solve
for the forces on gravitating particles and maps their masses onto the mesh for computing the dark matter gravitational interaction with the gas. The Poisson equation for the gravitational potential is computed
using a successive-overrelaxation solver. The solution of the MHD and Poisson equations is performed on the GPU; the handling of the mesh and particles is performed on the CPU.}

\textbf{GAMER-2 solves these equations using an adaptive mesh refinement scheme, which partitions the mesh into sub-grids of various sizes throughout the simulation domain such that higher resolutions (smaller cell sizes) are only used where needed, such
as in the high-density cores of clusters and at the gas discontinuities
formed in cluster mergers such as shocks and cold fronts. The refinement criteria employed are 1) the ratio of the second and first derivatives, known as the L\"ohner error estimator \citep{Lohner1987}, for gas density, pressure, temperature, which captures discontinuities such as shocks and cold fronts, and 2) grids--as well as their 26 nearest neighbours--are refined if they contain over 100 particles.} The simulations are run in boxes of $(14~\rm Mpc)^3$, split first into 128 cells per side \textbf{for the ``root'' grid} and then adaptively refined up to 4 times, yielding a maximum resolution of 6.8 kpc.

\begin{figure}
    \centering
    \includegraphics[width=0.45\textwidth]{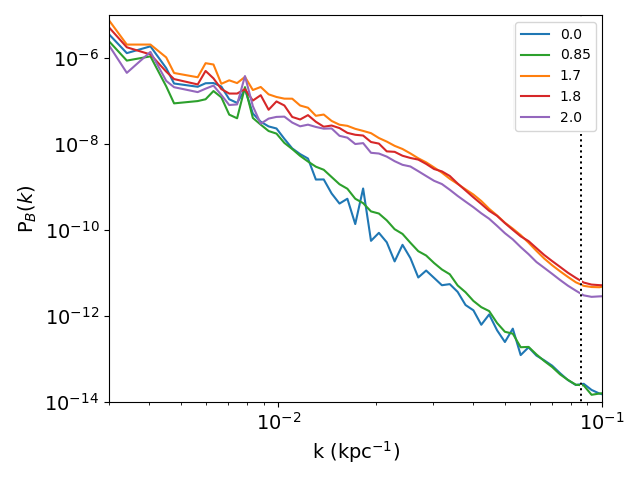}
    \caption{Power spectrum of the magnetic field strength for the $\beta_i=50$ run over time. The dotted line shows the Nyquist limit $1/2\delta x$. The magnetic field is initialised with a Kolmogorov spectrum (t = 0 Gyr, blue). It does not change much by 0.85 Gyr, as the clusters move towards each other. The power rises significantly at scales below 100~kpc at pericenter passage (t = 1.7 Gyr, orange), and decreases slowly for at least 0.3~Gyr after.}
    \label{fig:Pk_B}
\end{figure}
We initialise two cluster halos with dark matter and non-radiative gas. The halos are described by the super-NFW profile \citep{Lilley2018}:
\begin{align}
    \rho(r) &= \frac{3M}{16\pi a^3}\frac{1}{(r/a)\times(1+r/a)^{5/2}} ,
\end{align}
where $M$ is the total mass of the halo, and the scale radius $a$ relates to the half-mass radius as $a = R_e/5.478$. The major advantage of this over the more conventional NFW profile \citep{Navarro1997} is that the total mass converges as $r \rightarrow \infty$. For a given total mass of the halo, we assign a fraction $f_{\rm gas} = 0.17$ to the gas mass, and the remainder to the dark matter profile. The primary halo has a virial mass of $5\times 10^{14}M_\odot$ and the secondary is $1.6\times10^{14}M_\odot$. The best fit dark matter concentration for the primary halo was found to be $c_1 = 5$ in Paper I. The concentration of the secondary halo could not be constrained \textbf{because the projected mass density of the subcluster core is so low compared to that of the total system, i.e. it is dynamically insignificant. Therefore, we are confident that this would not affect the evolution of the magnetic fields, and }set it also to $c_2 = 5$. The impact parameter of the merger was constrained to be around 100~kpc, and the relative velocity 1200 km/s. 

The gas density profile is modeled using the formulation of \citet{Vikhlinin2006}:
\begin{align}
 n_pn_e &= n_0^2\frac{\left(r/r_c\right)^{-\alpha}}{\left(1+r^2/r_c^2\right)^{3\beta - \alpha/2}} \frac{1}{\left(1+r^\gamma/r_s^\gamma\right)^{\epsilon/\gamma}},
 \end{align}
where the normalisation $n_0$ is set so that the gas fraction within $R_{200}$ matches the universal average of 0.17. Through a parameter exploration described in more detail in Paper I, we found that the subclusters in Abell 2146 both initially had cool cores, and are well described by $\alpha = 2, r_s = 0.6r_{\rm vir}, r_c = 0.1r_{\rm vir}, \beta = 2/3, \gamma = 3$ and $ \epsilon = 3$.  

\begin{figure*}
    \centering
    \includegraphics[width=\textwidth]{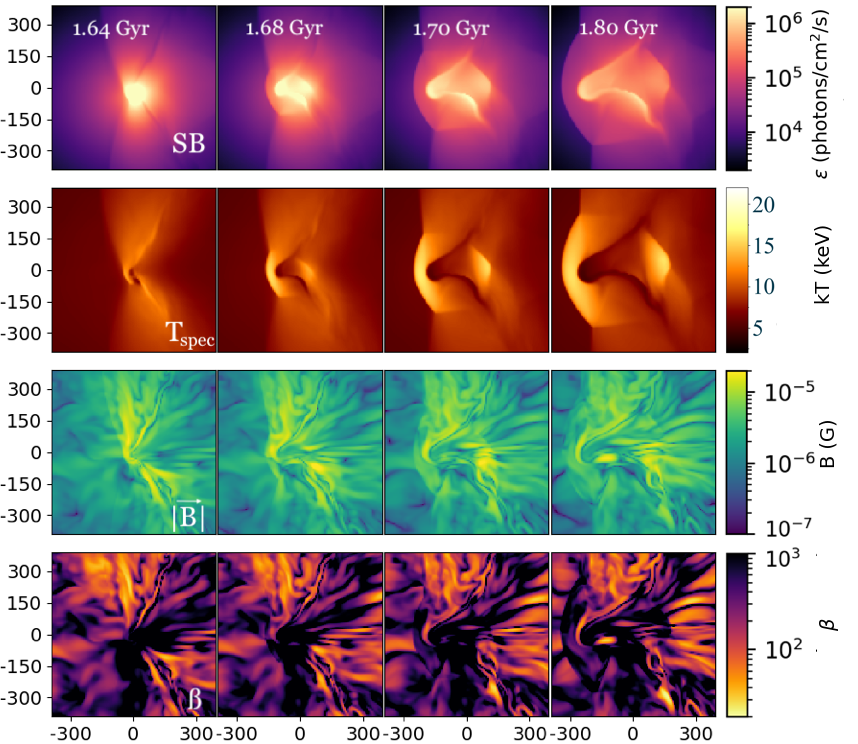}
    \caption{Projected X-ray surface brightness (photons/cm$^2$/s, top), spectral-weighted temperature (keV, second), slice of magnetic field strength $|\vec{B}|$ (G, third) and slice of $\beta = P_{\rm th}/P_{B}$ (dimensionless, bottom row) around pericentre passage for an initial average $\beta_i = 200$. Slices have a width of $\Delta x$ = 6.8~kpc and are taken in the plane of the merger, i.e., in the plane including both their potential minima and perpendicular to their angular momentum vector. The contact discontinuity is seen as a low surface brightness, high temperature, V-shaped feature initially ahead of the subcluster core, but connecting with it by pericentre passage at $t = 1.50$~Gyr. The magnetic field gets most amplified in the wake of the secondary subcluster, on the side closer to the core of the primary cluster. Here, $\beta$ is of order 10, so the magnetic field is dynamically significant. Also in the wake of the subcluster there are ripples, which look like KHI.}
    \label{fig:evolution}
\end{figure*}

\begin{figure*}
    \centering
    \includegraphics[width=0.9\textwidth]{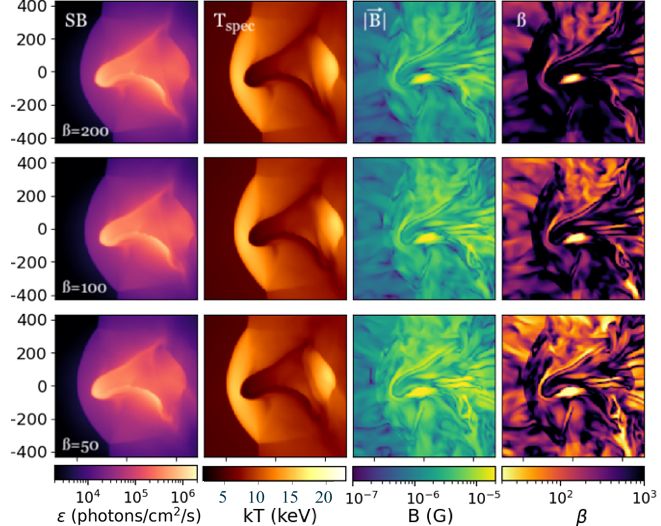}
    \caption{ Comparison of emission-weighted surface brightness (left column), spectral-weighted temperature (second column), slice of magnetic field strength (third column) and slice of $\beta$ (right column) for $\beta_i = 200$ (top row), $\beta_i=100$ (middle row) and $\beta_i = 50$ (bottom row). The colorbars are identical to Fig.~\ref{fig:evolution}. The ripples in the SB and temperature maps correspond to regions of ripples in the magnetic field. They are more prominent for stronger initial magnetic fields, because while magnetic tension can slow the growth of instabilities, in this case, the randomness of the seed field is also the only source of perturbations in the otherwise smooth cluster gas.}
    \label{fig:allbeta}
\end{figure*}

The magnetic field setup is described in more detail in \citet[][and references therein]{Brzycki2019}; here, we provide a brief summary. The field is initialised to be tangled (i.e., randomly oriented) with constant-$\beta$, meaning that in every radial aperture, the ratio between the magnetic and thermal pressures is roughly constant. Note that this is different from the gas density slope $\beta$, which we will no longer refer to in this work. The tangled field has a Kolmogorov power spectrum, $E(k) \propto k^{-5/3}$ with low and high scale cutoffs at 10 and 1000 kpc, respectively. The initial field is ``cleaned'' so as to remove any divergence, making $\nabla \cdot\mathbf{B} = 0$. We consider magnetic field strengths corresponding to $\beta$ = 200, 100 and 50; note that higher values mean weaker fields. 

Since the initial magnetic field is random, generating a fresh set for every $\beta$ would produce different realisations of a random distribution. We want to isolate the effect of increasing the magnetic field strength, not to confound it with slightly different initial distributions. Therefore, we only generated a random field for the weakest case, $\beta = 200$. Since $\beta \propto 1/P_B \propto B^{-2}$, we multiplied the field strength by $\sqrt{2}$ for $\beta = 100$ and by 2 for $\beta = 50$. \textbf{We also re-ran the $\beta=100$ simulation with three additional levels of refinement to investigate resolution effects, see \S \ref{sec:restest}.}

Paper I concluded that the merger was fairly close to the plane of the sky, $\theta \sim 30^\circ$; for convenience, we present most of our results projected along the z-axis, i.e., onto the plane of the merger. The projected temperature map is computed using the spectroscopic-like weighting $w \propto \rho^2T^{-3/4}$ \citep{Mazzotta2004}.

\section{Results}
\label{sec:results}

\begin{figure*}
    \includegraphics[width=\textwidth]{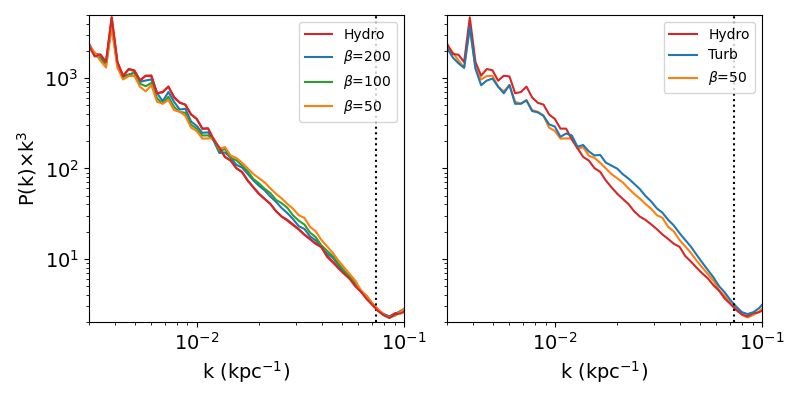}
    \caption{ \textit{Left panel:} Power spectra of the velocity as a function of the scale $k = 1/l_{\rm kpc}$ in the hydrodynamic and three MHD simulations in the inertial regime, as determined from the t = 1.80 Gyr line in Fig \ref{fig:Pk_B}. \textit{Right panel:} Power spectra of the velocity for $\beta=50$ MHD (orange), seed $\beta=50$ with $\vec{B}$ switched off at 0.85~Gyr to simulate turbulence (green) and the hydrodynamic case (red). Both power spectra are scaled by $k^4$ to highlight the differences. The power in gas motions increases in this regime as the magnetic field gets stronger ($\beta$ decreases). The power in the hydrodynamic case is lower than that of the MHD runs, while the power for turbulence without magnetic fields is higher. The bumps at high k (low scales) are due to aliasing, a feature of Fourier transforms in boxes with finite resolution; these appear more prominent due to the $k^3$ scaling.}
    \label{fig:Pk_allbeta}
\end{figure*}

\subsection{Evolution of Magnetized ICM during Cluster Merger}

Fig.~\ref{fig:evolution} shows the evolution of the central 1 Mpc of the merging system for $\beta_i = 200$ in the 0.12~Gyr surrounding pericentre passage. The top row shows projections of the surface brightness, the second row projections of the spectral-weighted temperature, the third row slices of the magnetic field strength and the bottom row slices of $\beta$. The evolution of the gas is described in detail in \S 3.2 of Paper I; here we provide a brief summary as relevant to the evolution of magnetic fields. As the subcluster falls into the potential of the primary halo from the right, the low density, high entropy gas from the outskirts of the two subclusters forms a contact discontinuity, an interface between the gas associated with each subcluster; the gas near this discontinuity is the first to be compressed during the merger. The core of the subcluster moves faster than the contact discontinuity, and the gas it displaces along its path moves out along the discontinuity. Pressure is almost continuous across the discontinuity, so the pressure gradient is the same on both sides, but the density differs, so gas on the low density side is accelerated to high speed in the direction parallel to the interface, creating a strong shear along it. This shear powers the growth of velocity fluctuations, initially created by Alfven waves from the tangled magnetic field, through the KHI. The compression and KHI also amplify the magnetic field along the interface, decreasing the plasma $\beta$ from initial values of 200 to $<$ 20. The bow shock forms close to pericentre passage and quickly overtakes the leftward travelling shock that was initially launched from the contact discontinuity. Meanwhile, the low-entropy subcluster core forms a second contact discontinuity, referred to henceforth as the merger cold front. It is around this cold front that the magnetic field gets the most amplified, as seen in the third row. Ram pressure stripping sweeps gas from the subcluster core, creating an obstruction to the sub cluster gas in its wake, which gives rise to the upstream shock. This does not correspond to a particularly strong feature in the magnetic field structure, but most of the field amplification occurs in the region between the cold front and the upstream shock, where the gas is colder and denser. However, the $\beta$ here is not as low as the $\beta$ value along the initial discontinuity, since the thermal pressure is also rather large. The key morphology of Abell 2146 is reproduced 0.10~Gyr after pericentre passage.

\begin{figure}
    \centering
    \includegraphics[width=0.5\textwidth]{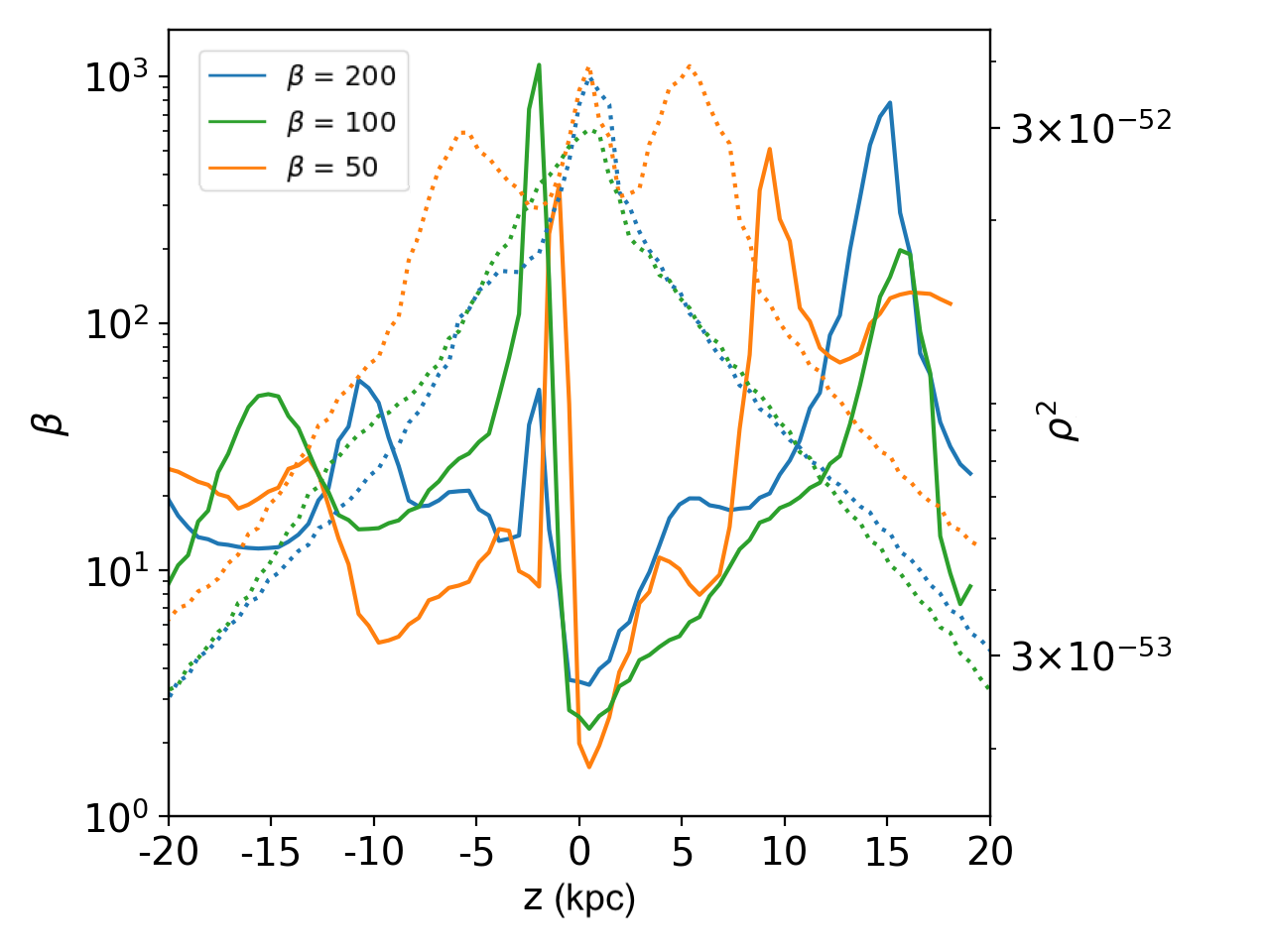}
    \caption{The ratio of thermal to magnetic pressure $\beta$ along the z-axis, with z=0 indicating the plane of the merger. The dotted lines show the square of the gas density, $\rho^2_{g}$, at the same positions. The region of low $\beta$, where the magnetic field is capable of dynamically affecting the gas, is just a sheet of thickness $\sim$ 1~kpc. The density in the surrounding regions, where 10 $< \beta < $ 1000, varies by less than an order of magnitude, so that they contribute more to the rotation measure in projection.}
    \label{fig:betaz}
\end{figure}

These simulations show that merging clusters need to be seeded with a realistic level of turbulence in order for KHI to produce potentially observable structure in the wake of the secondary subcluster. Conversely, this means that the observed structure of the wake provides an opportunity to constrain the level of turbulence in the merging subclusters. MHD effects on these features, although present, will be difficult to separate from the level of seeded turbulence - at least in the absence of other data to constrain the strength of the magnetic field. In cosmological simulations, this can likely be achieved self-consistently, since primordial magnetic fields are frozen into the plasma well before they start collapsing into halos, get amplified through turbulence associated with collapse, accretion and mergers, and end up with a random, Kolmogorov-like spectrum; this is indeed seen in past studies \citep[e.g.,][]{Dolag2005,Vazza2009, Vazza2011}.


\begin{figure*}
    \centering
    \includegraphics[width=\textwidth]{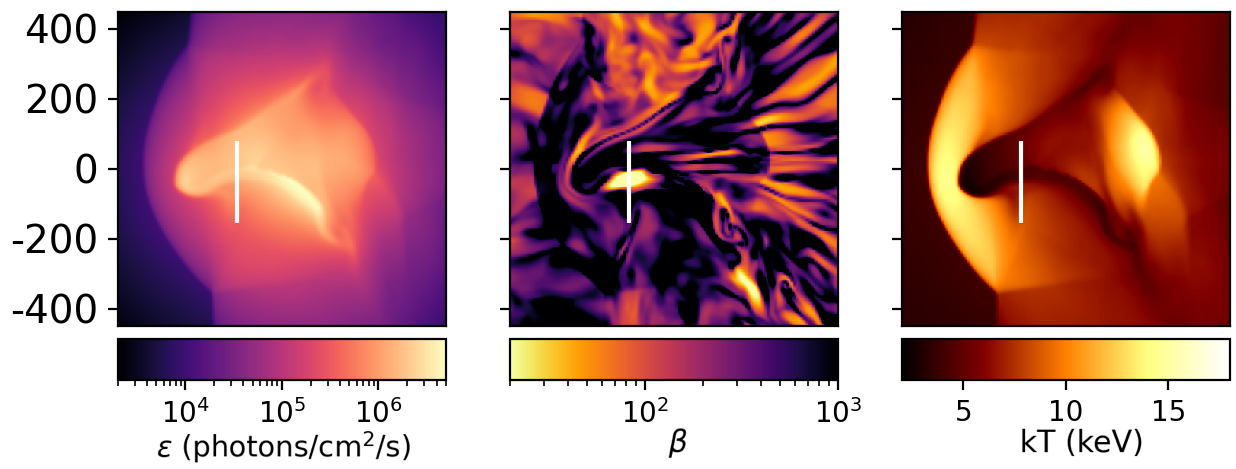}
    \caption{X-ray surface brightness (left), projected $\beta$ (middle), and spectral-weighted temperature (right) maps for the simulation with $\beta_i = 100$. The vertical lines were chosen to pass through the region of lowest $\beta$, where the magnetic field is most dynamically significant.}
    \label{fig:fingers}
\end{figure*}

\subsection{KHI ripples behind the subcluster}
One obvious effect of adding magnetic fields is the appearance of KHI-like ripples in the wake of the infalling cluster. We see that the interface associated with this instability is the contact discontinuity that forms in the early stages of the merger, which appears to extend along the edges of the bullet-like cold front in its wake. The gas in this discontinuity is pushed upstream of the subcluster core and away from the merger axis, creating a shearing layer. Perturbations in density, velocity, pressure, etc. along this shearing layer are expected to grow by the KHI. Fig.~\ref{fig:allbeta} shows the surface brightness, temperature, magnetic field strength and plasma $\beta$ for simulations with different seed field strengths, and confirms that the ripples are more prominent for stronger magnetic fields. While the field strength and $\beta$ show that the amplitude of the KHI ripples increases along the discontinuity with distance from the cool core, they are more visible closer to the core where the gas is denser and more luminous.

We quantify the relative prominence of the KHI ripples using the velocity power spectrum. We interpolated each component of the velocity ($v_x, v_y, v_z$) of the gas within a cube of side 0.5 Mpc, centreed on the potential minimum, onto a uniform grid of size $(256)^3$. The velocity grid was filtered with a Hamming window function to mitigate boundary effects in the Fast Fourier Transform (FFT). The FFT then yielded a 3D power spectrum, which was then binned into a 1D power spectrum for that particular velocity component. We note that a significant contribution to the velocity spectrum comes from the bulk motions of the merger; since the merger configurations are identical, the differences between the power spectra can be ascribed to the different strengths of the magnetic field in the different runs. The left panel of Fig.~\ref{fig:Pk_allbeta} shows the power spectrum of the total velocity, $P(k) = P_x(k) + P_y(k) + P_z(k)$, multiplied by $k^3$ to highlight the differences. The dotted line shows the Nyquist limit, $1/2\Delta x$. The magnitude of the power spectrum increases with decreasing $\beta$, i.e., it increases along with the strength of the magnetic field. \textbf{The effect is marginal, however, with the slope $\alpha$ of the power spectrum ($P (k) \propto k^\alpha$) going from -5.320 in the hydrodynamic case to -5.325 for $\beta_i = 50$.}
\begin{figure}
    \centering
    \includegraphics[height=0.45\textwidth]{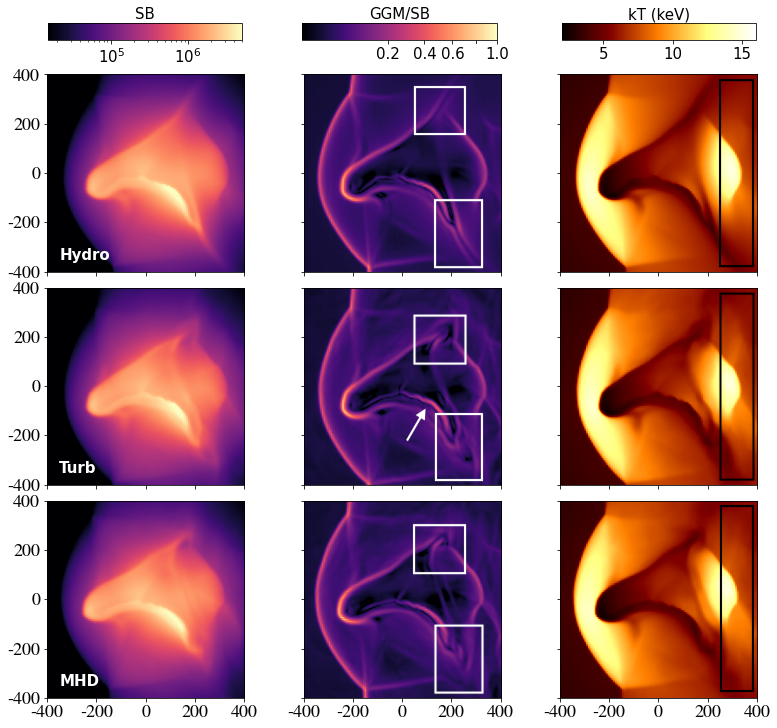}
    \caption{ X-ray surface brightness (left), Gaussian Gradient Magnitude (GGM) of the surface brightness (middle) and temperature (right) for the simulations with only hydrodynamics (top), seed $\beta = 50$ to generate turbulence, with the magnetic fields turned off at 0.85~Gyr (middle) and MHD with $\beta = 50$ throughout (bottom). The middle row shows what would happen if there were turbulent fluctuations equivalent to those in the bottom row, but there were no magnetic fields shaping their growth during the merger. The GGM highlights sharp features in the surface brightness image, making it easier to identify the ripples in the subcluster wake, highlighted in the white boxes. The arrow shows the additional KHI in the bridge connecting the two cool cores in the turbulent case, which is damped if magnetic fields are kept on. The black box in the temperature maps shows a larger-scale fluctuation in the brightness behind the upstream shock in the turbulent case, which also is damped if MHD is kept on and does not appear in the hydrodynamic case.}
    \label{fig:MHDOffTest}
\end{figure}

Our result, that the amplitude of KHI modes is greater with a stronger magnetic field, is contrary to earlier studies using uniform magnetic fields \citep[e.g.,][]{Dursi2008, ZuHone2011}. On the other hand, simulations of a rigid body in a turbulent magnetic field did find an amplified, turbulent magnetic wake forming behind the perturber \citep[][]{Asai2007, Takizawa2008, Ruszkowski2007}. When the magnetic field is uniform, we isolate the effect of magnetic tension, which resists the bending of a fluid element into eddies. Random magnetic fields, on the other hand, generate the seed velocity fluctuations themselves; the magnetic field lines want to be straight, but as they straighten, field energy is converted to kinetic energy, so there is overshoot and Alfven waves are produced in the cluster gas. Alfven waves produce shearing velocities, similar to unstable KHI modes; as such, they provide effective seeds for KHI. Figs \ref{fig:allbeta} and \ref{fig:Pk_allbeta} suggest that the effect of the stronger seed fluctuations outweighs the resistance to KHI due to magnetic tension in the draped field layer. 

To isolate the effect of turbulence from magnetic fields, we ran a simulation for $\beta_i = 50$ but turned the magnetic fields off at $t = 0.7$~Gyr. This gives enough time for the magnetic fields to generate velocity fluctuations in the cluster gas, but is well before the merger starts amplifying the magnetic field in the subcluster. Then we see what happens in the wake of the cluster, with the fluctuations seeded - but no longer affected - by magnetic fields. We chose the strongest seed field so that the effects are more visible. 

The resulting maps are shown in Fig.~\ref{fig:MHDOffTest}. The top panel shows the hydrodynamic simulation; the middle panel shows the MHD simulation with $\beta=50$, and the bottom panel started with the same magnetic fields as the middle panel, but these are switched off at 0.7~Gyr once they have seeded a random field of fluctuations analogous to turbulence. The left column shows the surface brightness. The right column shows the spectral-weighted projected temperature. The middle column takes the Gaussian Gradient Magnitude of the surface brightness \citep[][]{Walker2016}, which highlights sharp features and makes the ripples much more visible by eye. The velocity power spectra for the three simulations are shown in the right panel of Fig.~\ref{fig:Pk_allbeta}. \textbf{There is indeed more power in simulation with MHD seeded and then switched off, than when the magnetic fields are on throughout. The slope $\alpha$ steepens further to -5.340.}

The ripples are much less visible in the hydrodynamic case. As we suspected, however, they do show up in the bottom panel of Fig.~\ref{fig:MHDOffTest}, where the turbulent fluctuation spectrum is seeded but the magnetic field then turned off. Lastly, the ripples have already evolved to the point of dissipation in the bottom panel, whereas in the middle panel they are still growing in size. In other words, if the magnetic fields continue to act throughout the simulation, they do suppress the growth of the fluctuations. Correspondingly, the hydrodynamic simulation with turbulence seeded by magnetic field has the highest normalisation for the velocity power spectrum on $\sim$ 10 kpc scales, the hydrodynamic simulation with no seeded turbulence has the lowest, and the MHD run lies in between the two.

\begin{figure*}
    \includegraphics[width=\textwidth]{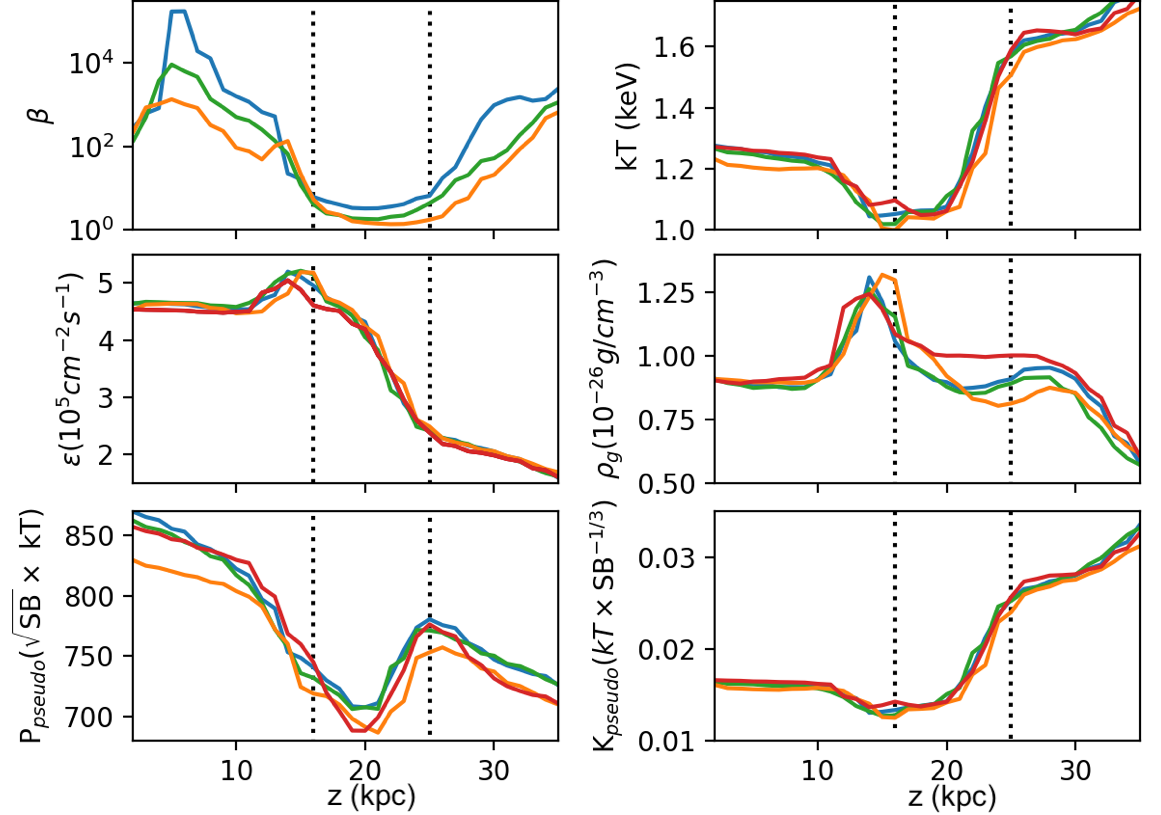} 
    \caption{Profiles for the plasma $\beta$, spectral-weighted projected temperature, photon emissivity, gas density, pseudo-pressure and pseudo-entropy, for $\beta$ = 200 (blue), 100 (green), 50 (orange) and the hydrodynamic case (red).The dotted lines mark the region where the magnetic field is the most amplified. The horizontal axis shows the position along the line region shown in Fig.~\ref{fig:fingers}. The $\beta$ profiles show that the field amplification saturates in this region, as also seen in the RM plots. The density in the plane of the merger is significantly lower in the presence of magnetic fields than without them; however, this effect is entirely erased in projection. Furthermore, the dip in surface brightness, and corresponding adiabatic increase in temperature, also occur without magnetic fields, so that detecting such a dim "channel" does not imply magnetic fields.}
    \label{fig:dipprof}
\end{figure*}

\subsection{Surface brightness channels}
\citet[][]{Werner2016}, in their $500 ks$ \textit{Chandra} observations and numerical simulations of sloshing cold fronts in the Virgo Cluster, found bands 10-15~kpc across with lower surface brightness that corresponded to regions of high magnetic field strength. The Virgo cluster is just 16.1 Mpc away from us, whereas Abell 2146 has an angular diameter distance almost 50 times greater. Therefore, a feature would have to be 50 times larger to be as well resolved in Abell 2146 as it is in Virgo. 
However, the scenario in Abell 2146 is also different - the cold fronts are associated with an ongoing merger, rather than sloshing; the much stronger bulk motions could, in principle, amplify magnetic fields a lot more. \citet[][]{Wang2016}, for example, claim to see such a channel in the merging cluster Abell 520. Here, the surface brightness dip is $\sim$400~kpc long, and aligns with the northern edge of the bridge behind the Bullet-like cool-core.

\begin{figure*}
    \centering
    \includegraphics[width=\textwidth]{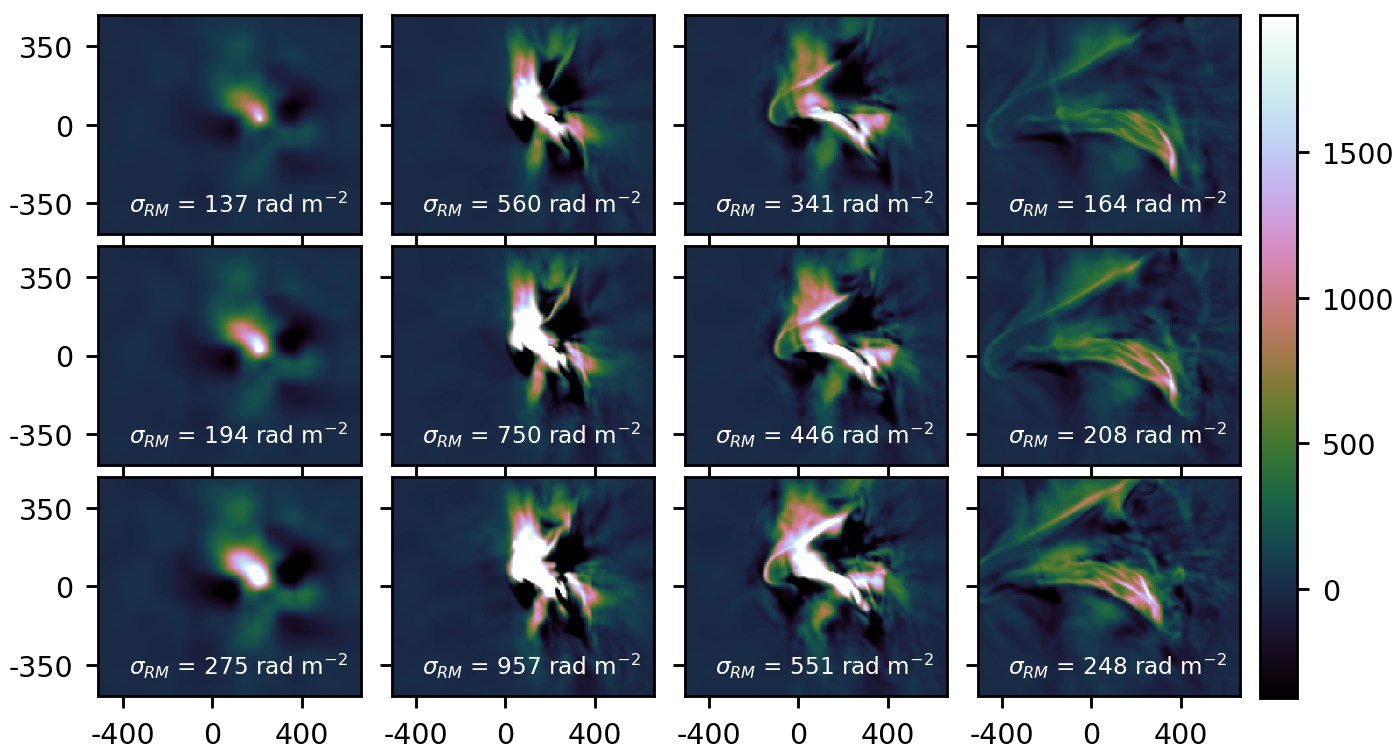}
    \caption{Faraday rotation measure (RM) in rad m$^{-2}$ for $\beta = 200$ (top), $\beta=100$ (middle) and $\beta=50$ (bottom). From left to right, the columns show the initial conditions, pericentre passage, and 0.1~Gyr and 0.3~Gyr after pericentre passage. \textbf{Each panel lists $\sigma_{RM} = \sqrt{\langle RM^2\rangle - \langle RM\rangle^2}$, computed here as the standard deviation of the RMs in every pixel of the image}. The magnetic field gets significantly amplified ($\sim4-5\times$) in the central 500 kpc right after pericentre passage. This amplification is temporary, and the field reaches  $2-3\times$ the initial value 0.1~Gyr post pericentre passage, the dynamical phase most consistent with observations of Abell 2146. The amplification is localised to the regions of large bulk motions, and remains close to 0 everywhere else.}
    \label{fig:rm}
\end{figure*}

Fig.~\ref{fig:fingers} shows the projected surface brightness, $\beta$ in the plane of the merger, and spectral-weighted projected temperature for the $\beta_i = 100$ simulation, with a vertical line in each passing through a region of lowest $\beta$. Fig.~\ref{fig:dipprof} then shows the 1D profile along the highlighted line for each of these quantities, as well as the density, pseudo-entropy $P = \sqrt{SB}\times k_BT$ and entropy index $K = k_BT\times SB^{-1/3}$.  In the distance range 15 - 25 kpc (marked with dotted vertical lines), where the magnetic field is the strongest, the temperature, density, pressure and photon emissivity all dip. The red line shows the hydrodynamic case, with no magnetic fields. The dip is just as visible in this case. The only property where the MHD profiles are significantly different is the slice of the density; this effect is washed out in projection, because, as shown in Fig \ref{fig:betaz}, the region of low $\beta$ is only a few kpc wide, and not much denser than the surrounding regions with much higher $\beta$. This suggests that the dim "channel" seen in Abell 520 can be explained hydrodynamically, without invoking the need for magnetic fields.

\subsection{Rotation Measure}
\begin{figure*}
    \centering
    \includegraphics[width=\textwidth]{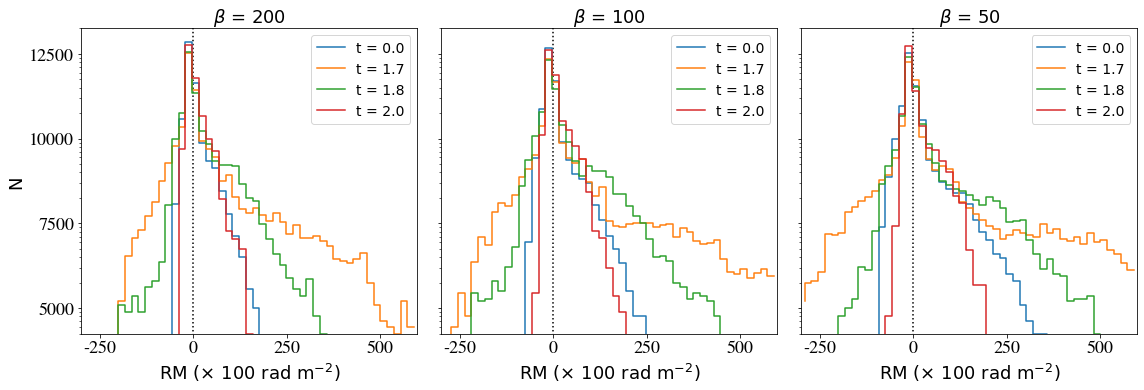}
    \caption{\textbf{PDF of the Faraday rotation measure in the central 250~kpc, where the field is most amplified by shearing motions, evolving over time in each simulation. As $\beta_i$ decreases and the seed field is stronger, the distribution broadens towards higher RM values. The field is most amplified at pericentre passage ($t = 1.70$~Gyr), as seen in the width of the RM distribution at this time. By $t=2.0$~Gyr, the RM distribution has returned to almost the initial configuration.}}
    \label{fig:rm-pdf}
\end{figure*}

Fig.~\ref{fig:rm} shows the Faraday rotation measure (RM) for hypothetical background radio sources for the MHD simulations with the magnetic field increasing downward, with $\beta =$200 (top), 100 (middle), and 50 (bottom), from the initial conditions (left column), through pericentre passage (2nd column), and 0.1 and 0.3~Gyr later on the right two columns, respectively. Again, the RM is proportional to the integral of the electron density times the component of the magnetic field parallel to the line of sight. \textbf{Fig \ref{fig:rm-pdf} summarizes these images by showing the probability distribution function (PDF) of the RM in the central 250~kpc; a narrower distribution implies RM is clustered towards smaller values.} 
If the merger is in the x-y plane, and we view it side-on, then this is $B_z$ integrated along the $\hat{z}$ axis. Since observations of RM are necessarily local, we also show the observable quantity $\sigma_{RM} = \sqrt{<RM^2> - <RM>^2}$, which is the standard deviation of the RM computed along different sight-lines; \textbf{in our case, each "sight-line" has the size of one resolution element (6.8 kpc/side)} \citep[e.g.,][]{Bonafede2010, Bohringer2016}.

\begin{figure*}
    \centering
    \includegraphics[width=\textwidth]{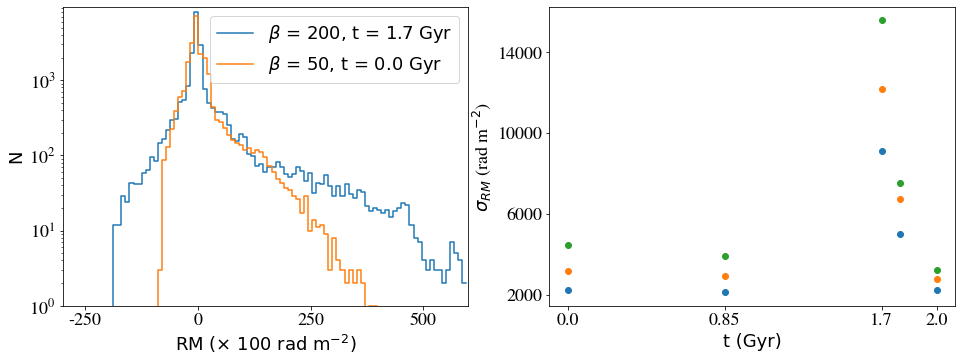}
    \caption{ \textit{Left:} The distribution of RM for a stronger seed field in a relaxed cluster (orange) and a weaker seed field, boosted at pericenter passage (blue). While the central regions may look similar, RM values greater than $\sim$100 rad$m^{-2}$ are highly unlikely without kinetic amplification. The absolute scale of this distinction will vary with the cluster mass in our construction, where the field strength scales with thermal pressure. \textit{Right:} The standard deviation of the rotation measure for $\beta = 200$ (blue), 100 (orange) and 50 (green) as a function of time. This highlights that the field is boosted by a factor of $\sim$4 at pericenter passage, but very briefly. It returns to the pre-merger state within 0.3 Gyr.}
    \label{fig:rm_t}
\end{figure*}

The initial values of $\sigma_{RM}$ scale linearly with the seed field, as expected. The kinetic energy of bulk motions amplifies the magnetic fields, so that the RM signal peaks at pericentre passage at $\sim3-4\times$ the initial value. The extent of field amplification is lower for stronger seed fields. This is the phenomenon of saturation - field amplification occurs when kinetic motions stretch magnetic field lines out; however, at some point, the magnetic stresses are so high that kinetic motions are unable to stretch the field lines any further. The amplification is, however, transient. Just 0.1~Gyr after pericentre passage, $\sigma_{RM}$ in this region has fallen by 40$\%$, most of the amplification at the interface of high shear velocity on the top and bottom edges of the bridge connecting the cool core remnants. Within another 0.2~Gyr, $\sigma_{RM}$ is almost the same as before the merger. \textbf{In terms of the PDF, this is seen as the distribution widening maximally at pericentre passage (t = 1.7 Gyr) and then reverting to the initial shape 0.3 Gyr later (t = 2.0 Gyr). This also means that if the RM is sampled in some finite number of sight-lines, this sample of values will have a higher variance for lower $\beta_i$ and closer to pericentre passage.} 
\textbf{Another thing worth noting is that the boosting of RM around pericentre passage is very different from a stronger seed field (see Fig \ref{fig:rm_t}). }

\begin{figure*}
    \centering
    \includegraphics[width=\textwidth]{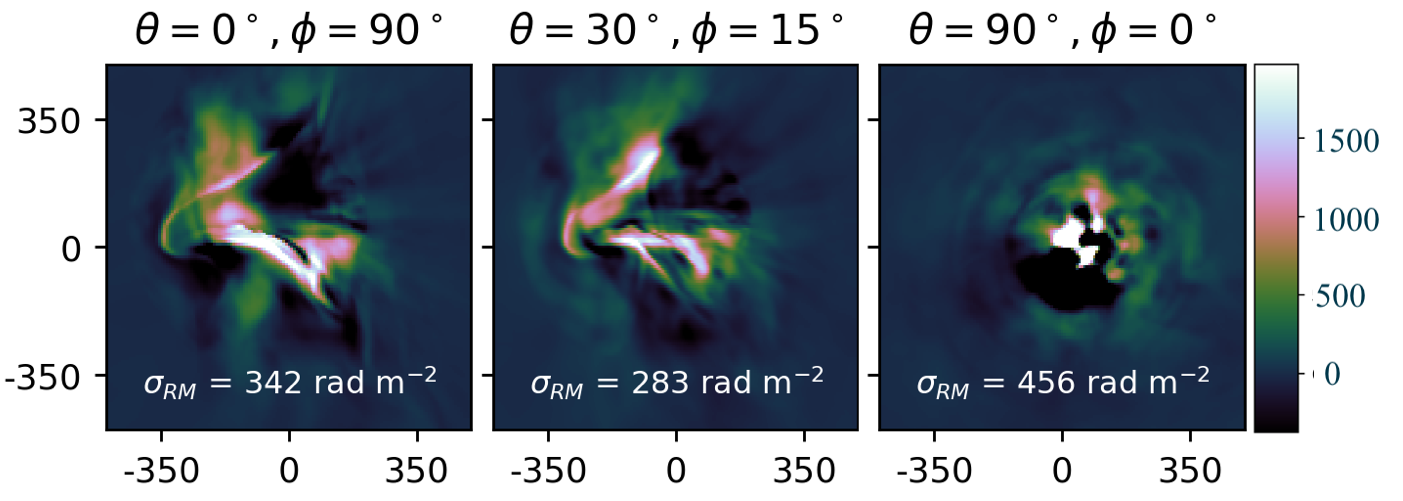}
    \caption{The effect of viewing direction $(\theta, \phi$) on the Faraday rotation measure (RM). The colorbar is identical to Fig \ref{fig:rm}. Following convention, $\theta$ is the angle between the observer and the y-z plane, and $\phi$ between the observer and the x-z plane. These plots are for the same simulation as the top row in Fig.~\ref{fig:rm}, at the snapshot in the third column. For $\theta=0$, changing $\phi$ has no effect on RM. The middle panel shows ($\theta=30^\circ, \phi=15^\circ$), the viewing direction most compatible with observations of Abell 2146, the RM varies by less than 18$\%$. If you look perfectly down the barrel (right), i.e., along the merger axis, RM is boosted by 33$\%$; this is largely due to the greater projected density.}
    \label{fig:rmrot}
\end{figure*}

Interestingly, even though the shearing motions are preferentially in the $x$-direction, corresponding to the initial velocity of the two halo centres, the field amplification is nearly isotropic. The RMS values of the $x$, $y$ and $z$ components of the magnetic field all grow by a factor of 1.5-2. This suggests that the average value of $\sigma_{RM}$ within some reasonable aperture, even as small as 100~kpc radius, should be relatively insensitive to the viewing direction. Fig.~\ref{fig:rmrot} shows the RM map for $\beta_i = 100$, $t = 1.80$~Gyr, for three different viewing directions within the range found consistent with the observed shock profiles and line-of-sight velocities in Paper I. Even setting $\theta=30^\circ$ reduces the measured $\sigma_{RM}$ by less than 18$\%$. 

Since RM measurements are done on small scales corresponding to the angular sizes of background radio sources, they should be interpreted with extreme caution around merging cluster cores. Fig.~\ref{fig:rm} also shows, however, that outside the central 100 kpc, the plasma $\beta$ remains similar to its initial value. Anomalously high RM measures are therefore an indicator of strong shearing motion. Conversely, if the goal is to measure the magnetic fields in relaxed clusters, a safe option might be to exclude the central region including any core remnants.

\subsection{Resolution effects}
\label{sec:restest}
\begin{figure*}
    \centering
    \includegraphics[width=\textwidth]{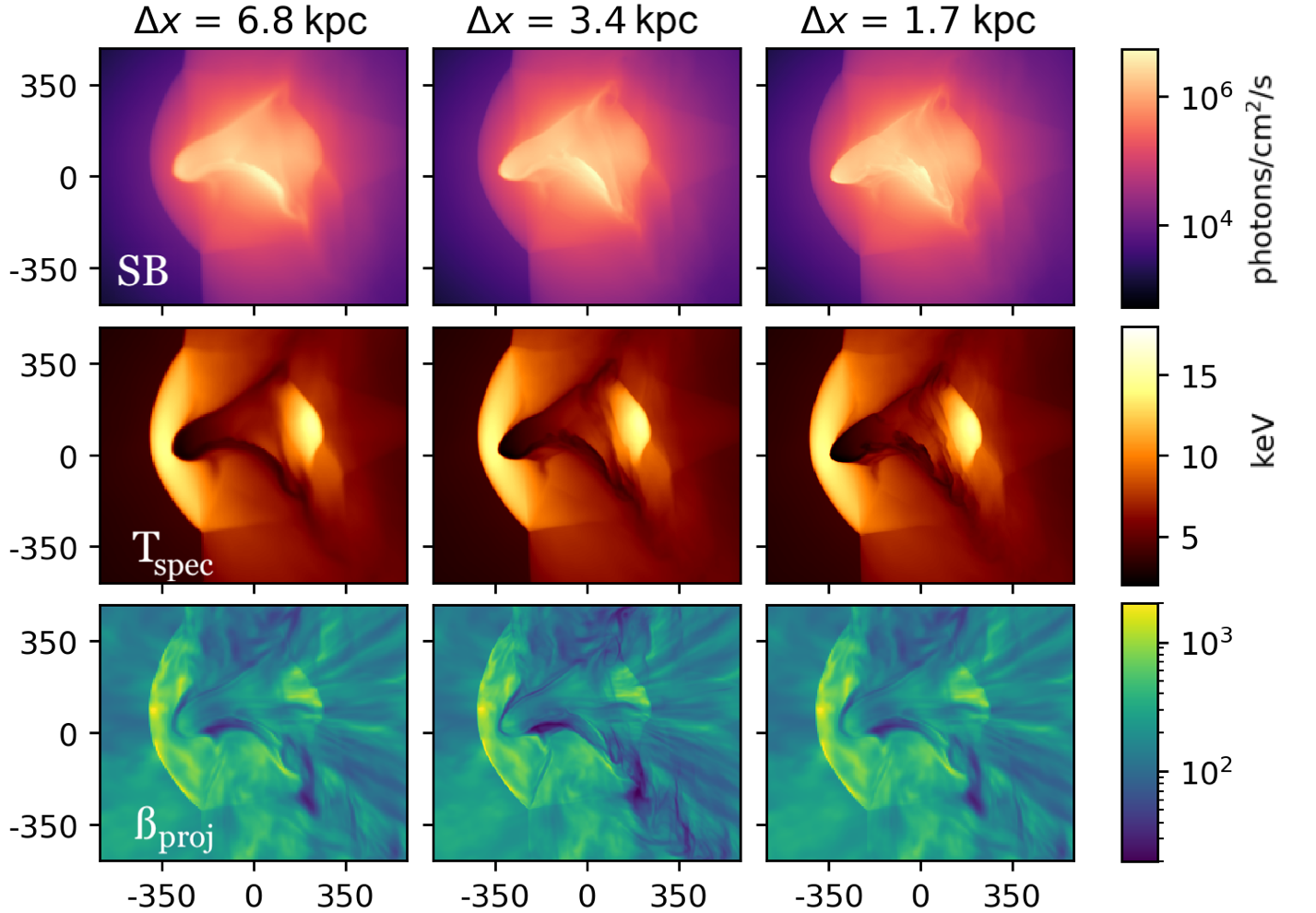}
    \caption{The effect of resolution on the surface brightness (top), spectral-weighted temperature (middle) and projected $\beta$ (bottom) of the system in the presence of magnetic fields with $\beta_i$=100. The default resolution of 6.8~kpc is shown on the left; the middle column sharpens the resolution to 3.4~kpc, and the right panel sharpens it again to 1.7~kpc. $\beta_{\rm proj}$ is the ratio of the projections of the thermal and magnetic pressures, respectively, each weighted by the square of the gas density, which affects their emissivity and therefore visibility. The number of levels of refinement goes from 4 on the left to 6 on the right, resulting in the specified resolutions. The simulations are shown at $t = 1.80$~Gyr, where the shock and cold front separations match observations. The higher resolution runs have lower numerical viscosity, and allow smaller-scale fluctuations to grow, resulting in more turbulent structure. Crucially, the additional features seen in temperature and surface brightness at higher resolution are not regions of low-$\beta$, suggesting that they are a result of turbulence rather than magnetic fields.}
    \label{fig:restest}
\end{figure*}

Fig.~\ref{fig:restest} shows the effect of increasing resolution on the evolution of the merger. All snapshots are at $t = 1.80$~Gyr, when the separations between the shock and cold fronts best match observations, and the initial average $\beta$=100. The top row shows the surface brightness at 0.3-7 keV, the second the Mazzotta-weighted temperature, and the bottom shows $\beta_{\rm proj}$, discussed below. From left to right, the resolution improves from 6.8 (our default spatial resolution) to 3.4 and 1.7~kpc. Observationally, we are limited by the \textit{Chandra} PSF of 1", which corresponds to 3.8~kpc at the redshift of Abell 2146. Therefore, the maps are all smoothed by a Gaussian of width 3.8~kpc.

In agreement with earlier studies \citep[e.g.,][]{ZuHone2011, ZuHone2015b}, greater resolution creates more turbulent structure in the temperature as well as surface brightness maps. Perturbations smaller than the simulation resolution get erased, a phenomenon called numerical viscosity; this applies both to seed fluctuations on small scales, and turbulence that cascades from higher to smaller scales. As a result, the KHI more efficiently grows eddies along the bridge between the cool core remnants, where the shear velocity is the highest. We note, however, that all the features in the higher-resolution runs also exist at lower-resolution, they are simply not as developed. This points further to the fact that the difference stems from reduced numerical viscosity.

The bottom panel of Fig \ref{fig:restest} shows the projected quantity $\beta_{\rm proj}$, the ratio of the projected thermal and magnetic pressures, each weighted by the square of the density, which decides their emissivity. We see that the new structures visible at higher resolution do not coincide with regions of low $\beta$, i.e., they are not due to the displacement of gas by magnetic pressure. Eddies in $\beta_{\rm proj}$ instead trace regions of the highest shear velocity, which indeed is what amplifies magnetic fields. It also grows KHI. In other words, regions of low $\beta$ also have KHI eddies, but the other way around is not necessarily true. Therefore, dips in the surface brightness and temperature maps are not necessarily evidence for the presence of magnetic fields.

\section{Discussion}
\label{sec:discuss}

\concsection{Reinterpreting previous studies of KHI in merging clusters}
Previous studies, such as \citet[][]{Vikhlinin2001}, use the absence of KHI ripples at the leading edge of a cold front as evidence for a strong magnetic field. This reasoning was based on theoretical work using uniform magnetic fields \citep[e.g.,][]{Dursi2008}. Instead, we highlight that magnetic fields are inherently turbulent in nature - they grow by turbulent amplification over the course of a cluster's history, and generate Alfven waves along the way. This point has already been made in both theoretical \citep[e.g.,][]{Subramanian2006,Donnert2018} and observational studies \citep[e.g.,][]{Dominguez2019, Stasyszyn2019}. In the context of merging clusters, therefore, we should neither treat the magnetic fields as uniform, nor forget that the turbulence that amplifies magnetic fields also seeds the instabilities which grow through the KHI. The level of turbulence is determined mainly by the recent growth history of a system. The amplitude of the KHI in a merging system might serve as a probe of the level of turbulence, particularly for high $\beta$, when the field has little direct impact on growth of KHI.

Second, a relatively strong magnetic field can suppress the growth of  KHI in these systems, but the effect is modest for realistic field strengths, which will make it hard to untangle such suppression from the impact of the initial level of turbulence. In our simulations, instabilities do not form at the leading edge of the cold front even when we seed turbulence using $\beta_i = 50$ and then turn magnetic fields off entirely, as seen in the bottom row of Fig.~\ref{fig:MHDOffTest}. Therefore, the absence of KHI at the leading edge of a cold front doesn't seem to require the presence of magnetic fields. Instead, it could be due to the lack of shear at the stagnation point, as described in \citet[][]{Churazov2004}.

Our result adds to an existing conversation about deviations from analytic models of fluid instabilities when we complicate the gas properties \citep[c.f.,][for reviews]{Donnert2018, Berlok2019}. The earliest analytic studies and MHD simulations modelled the core of the less massive cluster as a rigid object moving through a uniform density plasma, and predicted very long-lasting draping of the magnetic fields around the cold fronts \citep[e.g.][]{Dursi2008, Lyutikov2006}. Soon after, studies showed that more realistic, tangled fields were amplified much less efficiently than uniform ones; the draping layers they produced also didn't live as long as those of their uniform counterparts \citep[][]{O'Neill2009}. \citet[][]{Roediger2013b} showed that KHI could be suppressed by moderate levels viscosity, a claim that was further supported by \citet[][]{ZuHone2015b}.

\concsection{On the visibility of low-$\beta$ channels}
\citet[][]{Werner2016}, in their simulations of sloshing cold fronts in a Virgo-like system, find dim bands in regions of very low-$\beta$, where the large magnetic pressure displaces the gas and reduces the emissivity; a similar feature is possibly observed outside a cold front in Abell 2142 \citep[][]{Wang2018}. \citet[][]{Wang2016} claim to see a similar dim channel along one edge of the low-entropy bridge connecting the disrupted remnants of the two cluster cores in the merging system Abell 520. While we see a dim region outside the boundary of the cold front in Abell 2146, we find that this is also produced in the hydrodynamic case, and is not uniquely produced by magnetic fields. A strong magnetic field, highly amplified by shearing motions across the cold front, does create a slight dip in density in the plane of the merger compared to the hydrodynamic case; however, this effect is restricted to a very thin slice, and the effect is lost in projection. We suspect that the difference between systems like Abell 2146 and Abell 520 on the one hand, and Virgo on the other, is that the cold fronts in the latter are formed through sloshing, and reinforced over time through the periodic nature of the gas motions. Abell 2146 experienced pericentre passage just $\sim$0.1~Gyr ago; it may experience sloshing over the course of several more gigayears, and any existing channels may become detectable even in projection. However, this hypothesis needs further testing, and is beyond the scope of this work.

\concsection{The discerning power of Faraday Rotation}
\citet[][]{Johnson2020} shows that the Faraday RM can only constrain the magnetic field strength of a cosmologically simulated cluster to within a factor of 3, due to inhomogeneities in ICM density and the unknown scaling between $\beta$ and $P_{\rm th}.$ In our study, for example, we have assumed that the magnetic pressure is a constant fraction of the thermal pressure in the initial conditions, whereas the causal relationship is between magnetic and turbulent pressure; the constant relation between turbulent and thermal pressure is a simplifying assumption on our part, albeit with observational support \citep[][]{Govoni2004, Govoni2017}. Since the RM is always sampled locally, in small regions with background radio sources along the line of sight, any spatial variation in this $\beta - P_{\rm th}$ relation would be significant. 

\concsection{Beyond MHD} Our MHD simulations of the merger in Abell 2146 were conducted to look for observable impacts of magnetic fields, prior to more physically complete simulations to include key transport processes as well.  Both thermal conduction and viscosity may have significant impacts on a dynamical event like a merger \citep{ZuHone2013, ZuHone2015}. However, recent studies have shown that thermal conduction, both isotropic and anisotropic, is likely very strongly suppressed in weakly magnetised plasmas like the ICM \citep{Roberg2016}. Anisotropic conduction was originally found to cause instabilities such as the heat-flux driven buoyancy instability in cool cluster cores, but interestingly, this appears to be almost entirely countered by anisotropic viscosity \citep{Kunz2011b, Kunz2012, Latter2012} and turbulence \citep{Ruszkowski2010}. Our understanding of instabilities in turbulent, high-$\beta$, low-density plasmas like the ICM is still evolving, and the interactions between them are not fully understood \citep[e.g.][]{Schekochihin2009, Maier2009}. However, if their net effect were significant, galaxy clusters would look significantly different from how they are currently observed - for example, the magnetothermal instability (MTI) would require cluster outskirts to be isothermal, which they are not. As we continue to unravel the complex effects of plasma instabilities, MHD continues to provide helpful, first-order insights into the role of magnetic fields in the ICM.

\section{Conclusions}
\label{sec:conclusions}
We presented magnetohydrodynamic (MHD) simulations with various strengths of the magnetic field for the best-fit dynamical model for the galaxy cluster Abell 2146 obtained in Paper I. The simulations used a tangled initial magnetic field with a constant ratio of thermal to magnetic pressure $\beta$ as a function of cluster-centric radius. We produced maps of the Faraday rotation measure and X-ray observables to search for detectable consequences of such a magnetic field. We found that:

\begin{itemize}
    \item The merger strongly amplifies magnetic fields in the central $\sim$200 kpc through shearing motions, though only for a brief period of time $\lesssim$0.3~Gyr. This increases the measured Faraday RM by a factor of 4 if our viewing direction is perpendicular to the merger plane, and more if instead the merging clusters have a relative velocity along our line of sight. For the most likely viewing direction towards Abell 2146, this additional boost is $\sim18\%$. 
    
    \item In the "wake" of the subcluster, on the top and bottom edges of the "bridge" connecting the cool core remnants,  KHI-like ripples form in the MHD simulations. This is because a tangled magnetic field is a source of perturbations in the ICM, which are amplified in the presence of large shearing motions along the interface. The amplitude of the KHI in a merger is sensitive to the initial level of turbulence in the ICM, which also amplifies seed magnetic fields. As a result, the velocity power spectra have higher magnitudes for stronger magnetic fields. This is at odds with earlier observational studies, which suggest that the primary effect of magnetic fields is to stabilise discontinuities against instabilities.
    
    \item Inspired by work on a similar merging cluster, Abell 520, we searched for low surface brightness "channels" that could result from the displacement of gas by high magnetic pressure. Whereas we do find a surface brightness dip at a region of very low $\beta$, this does not correlate with $\beta$, and indeed is also produced in the hydrodynamic simulations. The gas in this region does get displaced in the plane of the merger, but the region of low $\beta$ is confined to a thin sheet, $\sim$1~kpc wide, which is not much denser than the surrounding regions of very high $\beta$, so that the effect is entirely lost in projection.

    \item The amplitude of the KHI depends on the resolution of the simulation. This is because a lower resolution imposes a larger numerical viscosity, damping or erasing fluctuations on smaller scales. In realistic galaxy clusters, we expect some non-zero plasma viscosity, although this magnitude is currently unconstrained. If this viscosity can be measured independently, we can pick a resolution with an equivalent numerical viscosity. This needs to be considered in any studies that aim to constrain plasma viscosity parameters or magnetic fields using the amplitudes of fluctuations in observable quantities. 
\end{itemize}

Our results emphasise the need to model turbulence along with magnetic fields to produce realistic galaxy clusters, and the fleeting but strong amplification of the Faraday rotation measure in the centres of merging clusters, even if viewed perpendicular to the direction of the merger velocity. Modelling the efficiency of transport processes, accounting for anisotropy in the presence of magnetic fields, is a promising next step in constraining the plasma microphysics of systems like Abell 2146.

\section*{Acknowledgements}
We thank the anonymous referee for the thoughtful discussion and constructive feedback. UC was supported by NASA grants G08-19108X and GO0-21070X. JAZ, PEJN and UC are funded in part by the Chandra X-ray centre, which is operated by the Smithsonian Astrophysical Observatory for and on behalf of NASA under contract NAS8-03060. The simulations were run on the Grace cluster at the Yale High-Performance Computing centre. 
\section*{Data Availability}

Snapshots of the simulations used for this project will sequentially be added to the Cluster Merger catalog \citep{ZuHone2018}, accessible at \url{http://gcmc.hub.yt/}. The full simulation files are too large to host permanently on a server, and will be shared on reasonable request to the corresponding author. 

\newcommand{\newblock}{}
\bibliographystyle{mnras}
\bibliography{reference.bib} 
\label{lastpage}
\end{document}